\documentclass{chjaa}                  
\def \th {\thinspace}

\def\approxgt{\mathrel{\hbox{\rlap{\lower.55ex \hbox {$\sim$}} \kern-.3em \raise.4ex \hbox{$>$}}}}
\def\lesssim{\mathrel{\hbox{\rlap{\lower.55ex \hbox {$\sim$}} \kern-.3em \raise.4ex \hbox{$<$}}}}
\def\approxlt{\mathrel{\hbox{\rlap{\lower.55ex \hbox {$\sim$}} \kern-.3em \raise.4ex \hbox{$<$}}}}

\usepackage{graphicx,times}             

\begin{document}

   \title{The importance of radiation pressure in the launching of jets
}

   \volnopage{Vol.0 (200x) No.0, 000--000}      
   \setcounter{page}{1}                         

   \author{M. J. Church
      \inst{1,2}\mailto{}
   \and N. K. Jackson
      \inst{1}
   \and M. Ba\l uci\'nska-Church
      \inst{1,2}
      }
   \institute{School of Physics and Astronomy, University of Birmingham, Birmingham B15 2TT, UK\\
             \email{mjc@star.sr.bham.ac.uk}
        \and
             Astronomical Observatory, Jagiellonian University, ul. Orla 171,
             30-244, Cracow, Poland\\
          }
   \date{Received~~2007 month day; accepted~~2007~~month day}

\abstract{
Based on the results of applying the extended ADC emission model to three Z-track sources:
GX\th 340+0, GX\th 5-1 and Cyg\th X-2, we propose an explanation of the Z-track sources
in which the Normal and Horizontal Branches are dominated by the increasing radiation pressure
of the neutron star. The emitted flux becomes several times super-Eddington at the Hard Apex
and Horizontal Branch and we suggest that the inner accretion disk is disrupted by this and 
that part of the accretion flow is diverted vertically. This position on the Z-track is exactly
the position where radio emission is detected showing the presence of jets. We thus propose
that high radiation pressure is a necessary condition for the launching of jets. We also show 
that flaring must consist of unstable nuclear burning and that the mass accretion rate 
per unit emitting area of the neutron star $\dot m$ at the onset of flaring agrees well 
with the critical theoretical value at which burning becomes unstable.
   \keywords{
   physical data and processes: acceleration of particles --- 
   physical data and processes: accretion: accretion disks ---
   stars: neutron --- stars: individual: \hbox{GX\th 340+0,} \hbox{GX\th 5-1,} \hbox{Cyg\th X-2} ---
   X-rays: binaries}
   }

   \authorrunning{M. J. Church, N. K. Jackson \& M. Ba\l uci\'nska-Church}            
   \titlerunning{Jet formation in the Z-track sources}                   

   \maketitle

%
\section{Introduction}           
\label{sect:intro}

The Z-track sources are the brightest group of Galactic low mass X-ray binaries (LMXB)
containing a neutron star persistently emitting at the Eddington luminosity or several times
this. The sources trace out a Z-shape in hardness-intensity (Hasinger et al. 1989)
clearly showing that strong
physical changes take place, probably at the inner disk and neutron star, but a convincing
explanation of the Z-track phenomenon does not exist. The majority of LMXB are of the Atoll class which show
somewhat different shapes in hardness-intensity which are also not understood, and neither
is the relation between the two classes making our understanding of LMXB very incomplete.

Moreover, it is well-known that the Z-track sources are detected as radio emitters, but
in one branch only, the horizontal branch. Not only is radio detected, but striking results
from the VLA show the release of a massive radio condensation from the source Sco\th X-1
(Fomalont et al. 2001). Because radio is detected essentially in one branch only, the
sources offer the possibility of determining
the conditions found in this branch distinguishing it from the other two branches, and so finding the 
conditions necessary for jet formation.


Possible ways of understanding the Z-track sources are by theoretical approaches, timing studies
or spectral studies. 
A theoretical model for the Z-track sources was produced by Psaltis et al. (1995) based on a
magnetosphere of the neutron star and the changing properties and geometry of this as the
mass accretion rate changed. However, the model assumed that the Comptonized emission observed 
in the spectra (of all LMXB) originated in a small central region close to the neutron star, and
this is inconsistent with our more recent measurements of Comptonizing region size  
(Church \& Ba\l uci\'nska-Church 2004, below). 

Extensive timing studies have been made to investigate 
QPO variations around the Z-track (e.g. van der Klis et al. 1987), but this has not revealed the 
nature of the Z-track. Previous spectral fitting has applied the Eastern model (Done et al. 2002,
Agrawal \& Sreekumar 2003; di Salvo et al. 2002) which assumes the X-ray emission consists of 
disk blackbody emission plus non-thermal emission from a small central Comptonizing region. 
However, our work over a period of 10 years with the dipping class of LMXB provides strong
evidence that the source of Comptonized emssion, the ADC, is very extended, typically having a 
radial extent that is 15\% of the accretion disk size, but increasing with source luminosity,
and this is inconsistent with the Eastern 
model. As a result we have proposed the ``extended ADC'' emission model consisting
of blackbody emission from the neutron star plus Comptonized emission from an extended ADC
(Church Ba\l uci\'nska-Church 1995). Moreover, the pattern of parameter changes obtained by 
fitting the Eastern model to the Z-track sources is not very easy to interpret and does 
not immediately suggest a 
convincing physical explanation. Thus in the present work, we take the approach of applying
the extended ADC model for the first time to the Z-track sources and we present
the results of applying this model to the sources GX\th 340+0, GX\th 5-1 and Cygnus\th X-2.

\begin{figure*}[!ht]
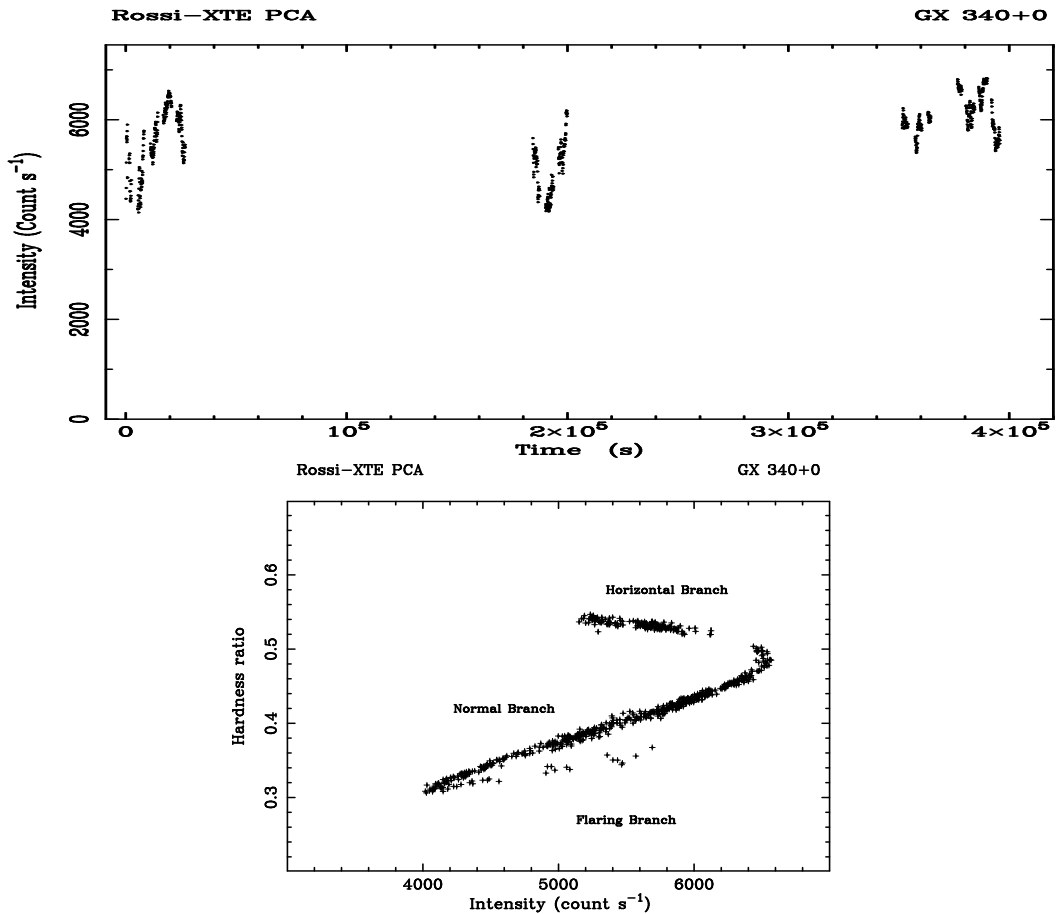
                                                           
\begin{center}
\includegraphics[width=60mm,height=140mm,angle=270]{church_2007_01_fig1a}   
\includegraphics[width=60mm,height=80mm,angle=270]{church_2007_01_fig1b} 
\caption{Top: Background-subtracted and deadtime-corrected PCA light curve of the 1997
September observation of GX\th 340+0 with 64 s binning. Bottom: the corresponding variation
of hardness ratio \hbox{(7.3 -- 18.1 keV)/(4.1 -- 7.3 keV)} with intensity.}
\label{}
\end{center}
\end{figure*}

\section{Observations and analysis}

We analysed the {\it Rossi-XTE} observation of GX\th 340+0 made on 1997 September lasting 400
ksec, the 1998 November observation of GX\th 5-1 spanning 95 ksec and the 235 ksec 1997 
June/July observation of 
Cygnus\th X-2. Data from both the proportional counter array (PCA: 2 - 60 keV) and the high energy
X-ray timing experiment (HEXTE: 15 - 250 keV) were used. Analysis was carried out using the standard
{\it RXTE} software {\sc ftools 5.3.1}. The background-subtracted, deadtime corrected PCA lightcurve
for GX\th 340+0 is shown in Fig. 1 (upper panel). A hardness ratio was defined as the ratio of the 
intensities in the bands 7.3 - 18.5 keV and 4.1 - 7.3 keV, and the hardness-intensity diagram is
shown in the lower panel of Fig. 1. Spectra were extracted corresponding to 10 positions about
equally spaced along the Z-track by defining for each narrow ranges of hardness ratio (0.01 wide) 
and intensity (100 c s$^{-1}$ wide). Good time interval files (GTI) for each selection
were also used to extract HEXTE spectra, and simultaneous fitting of the PCA and HEXTE data
at each position of the Z-track carried out applying the extended ADC model. Analysis of
the observations of GX\th 5-1 and Cyg\th X-2 were carried out in the same way, and results of
spectral fitting are given below.


\section{Results}

\begin{figure}
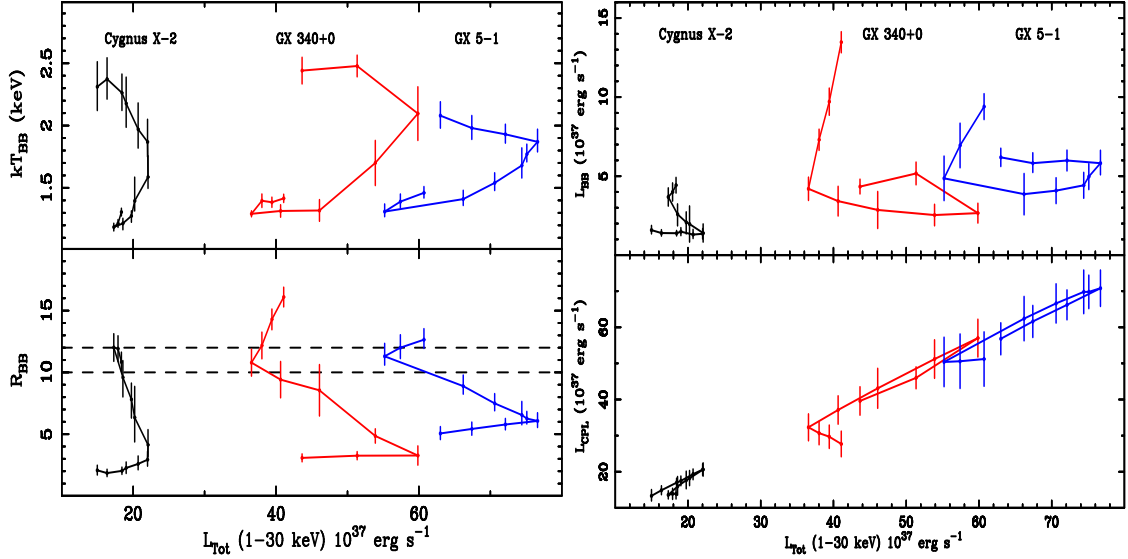
                                                             
\begin{center}
\includegraphics[width=74mm,height=74mm,angle=270]{church_2007_01_fig2a}                     
\includegraphics[width=74mm,height=74mm,angle=270]{church_2007_01_fig2b}         
\caption{Left: blackbody temperature (upper panel) and radius (lower panel) as a function of the
total luminosity; right: the individual luminosities of the neutron star blackbody (upper panel) 
and of the Comptonized ADC emission as a function of the total luminosity (lower panel).}
\end{center}
\end{figure}

The Z-track shown in Fig. 1 clearly shows the three branches: the horizontal branch (HB), normal 
branch (NB) and flaring branch (FB), these branches corresponding to particular sections of the
lightcurve: for example, the strong flaring in the early part of the observation provides the FB.
Based to some extent on a multi-wavelenth campaign on Cyg\th X-2 (Hasinger et al. 1990),
there has been a widely-held view that the changes taking place along the Z-track
are in some way driven by a mass accretion rate $\dot M$ that increases monotonically 
in the direction HB - NB - FB. We note at this stage that this appears inconsistent
with an X-ray intensity that {\it decreases} moving on the normal branch in this direction. 

The spectral fitting results were very robust partly as a result of the high-quality data
with typically 1 million counts in each spectrum. It was also clear that the use of the extended 
ADC model not only provided very good fits, but also results that could be easily interpreted
in a straightforward way. We present firstly the results for the neutron star blackbody emission 
which is described by the temperature $kT_{\rm BB}$ and the blackbody radius $R_{\rm BB}$ which 
provides the emitting area. Figure 2 (left) shows $kT_{\rm BB}$ for all three sources in the upper panel
and $R_{\rm BB}$ in the lower panel.

A clear pattern of systematic variation is evident. In all three sources the temperature is
lowest at $\sim$1.3 keV at the soft apex of the Z-track, i.e. the apex between the normal
branch and flaring branch, suggesting that the mass accretion rate
$\dot M$ is lowest. At this position $R_{\rm BB}$ is maximum and 10 - 12 km in all sources,
having a mean value of 11.4 $\pm$ 0.6 km at 90\% confidence suggesting that the whole neutron 
star is emitting, and on this assumption, the analysis provides a measurement technique for 
the neutron star radius. We thus propose that the soft apex is a quiescent state of the sources. 

Next in Fig. 2 (right), we show the individual luminosities of the neutron star blackbody 
and the Comptonized emission of the ADC for all three sources as a 
function of the total 1 - 30 keV luminosity. Concentrating on the
Comptonized emission in GX\th 340+0, it can be seen that this is the dominant emission component,
ten times more luminous than the blackbody, and that as the source moves up the Z-track from the
soft apex to the hard apex between the normal branch and horizontal branch,
this component doubles in luminosity. The X-ray intensity also, of course,
increases by a similar factor and we suggest therefore that the mass accretion rate is increasing
in this direction {\it contrary} to the widely-held view that $\dot M$ increases monotonically round the
Z-track in the direction HB - NB - FB (Priedhorsky et al. 1986). 
For further discussion of this point see Church et al. 
(2006). Our suggested increase of $\dot M$ is consistent with the observed increase of blackbody
temperature as more accretion reaches the surface of the neutron star. On the horizontal branch
the luminosity of the Comptonized emission falls almost back to its initial value. All three
sources exhibit the same behaviour and its is remarkable that all three sources lie on
the same line in the lower panel of Fig. 3 indicating that the luminosity of the ADC emission is simply the same
function of $\dot M$ (i.e. the total luminosity) in all sources. The blackbody luminosity
in the upper panel of Fig 3. does not vary greatly on the NB and HB because of the combined
effects of changes in $R_{\rm BB}$ and $kT_{\rm BB}$ but increases in flaring as discussed below.

\begin{figure*}[!h]
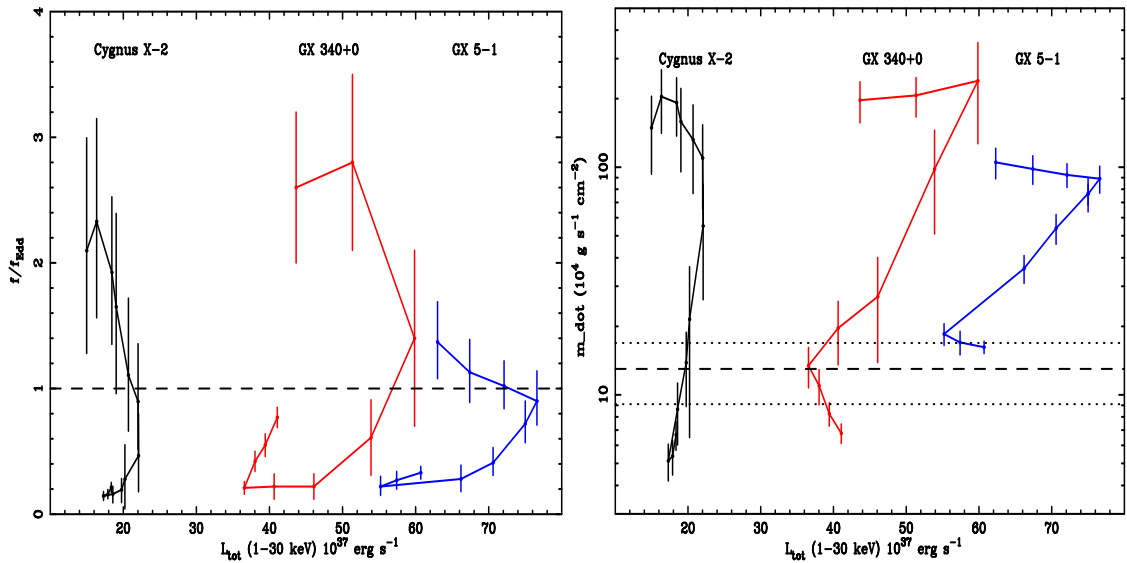
                                                                        
\begin{center}
\includegraphics[width=74mm,height=74mm,angle=270]{church_2007_01_fig3a}  
\includegraphics[width=74mm,height=74mm,angle=270]{church_2007_01_fig3b}  
\caption{Left: flux of the emitting part of the neutron star as a fraction of the Eddington flux
(see text), right: mass accretion rate per unit emitting area of the neutron star $\dot m$.}
\label{}
\end{center}
\end{figure*}

The blackbody temperature increase by a factor of two means that $T^4$ increases by nearly an
order of magnitude. Because of the decrease in $R_{\rm BB}$ as the source moves from the soft
to the hard apex, we consider the change in radiation pressure, not in terms of the Eddington
luminosity, but in terms of the emitted flux of the neutron star. At the hard apex the emitting
area is reduced to an equatorial belt on the neutron star and in Fig. 3 (left) we show the
emitted flux $f$ as a fraction of the Eddington flux $f_{\rm Edd}$ = $L_{\rm Edd}/ {4\,\pi\,R^2}$,
where $R$ is the radius of the neutron star assumed to be 10 km.
In all three sources this ratio rises from low values
at the soft apex $\sim$20\% to super-Eddington values at the hard apex and on the horizontal branch.
However, these are exactly the positions at which radio emission is detected indicating the
presence of jets, and so we propose that high radiation  pressure plays a major role in launching 
the jets, by disrupting the inner disk and diverting accretion flow into the vertical direction
(Sect. 4). In a more detailed discussion (Church et al. 2006), we show that the reduction
in blackbody radius is consistent with this disruption.


\begin{figure*}[!h]                                                                 
\begin{center}
\includegraphics[width=110mm,height=44mm,angle=0]{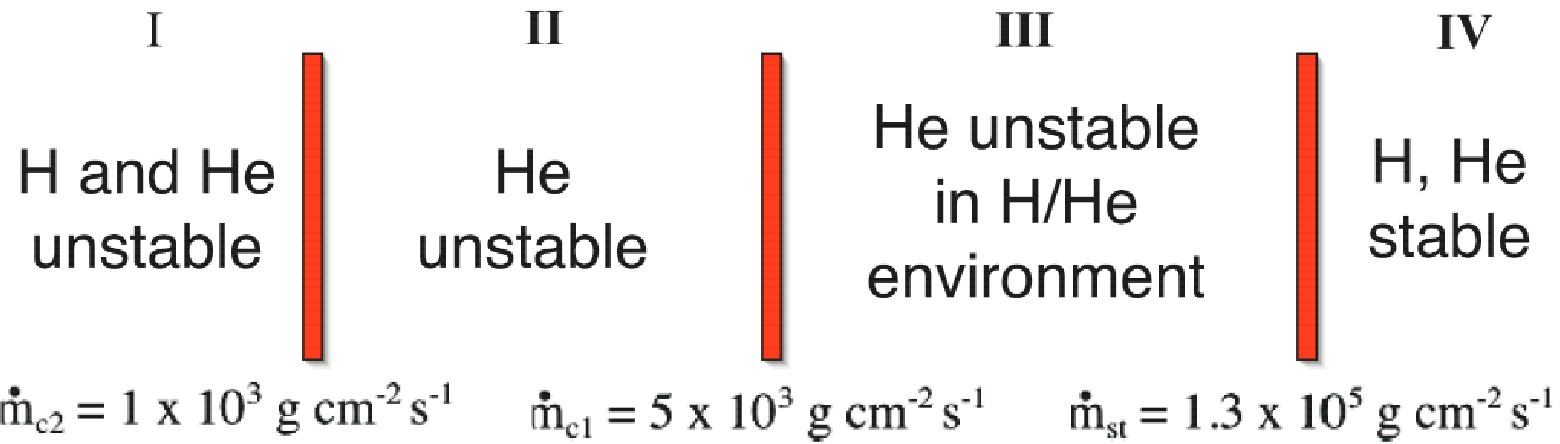}
\caption{The four r\'egimes of stable or unstable nuclear burning from Bildsten (1998),
showing the critical values of $\dot m$ that demarcate the r\'egimes.}
\label{}
\end{center}
\end{figure*}

In flaring, the luminosity of the Comptonized emission essentially does not change
(Fig. 2) so that $\dot M$ is constant
and we can conclude that the strong increase of intensity in flaring must be due to
unstable nuclear burning and cannot be due to an increase of $\dot M$, resolving a
longterm controversy on this issue. In the theory of unstable nuclear burning (Fujimoto et al. 1981; 
Fushiki \& Lamb 1987; Bildsten 1998; Schatz et al. 1999)
the physical conditions in the atmosphere of the neutron star depend on the mass accretion
rate per unit area $\dot m$, i.e. $\dot M$ divided by the emitting area. In Fig. 3 (right) we 
comapare $\dot m$ at the soft apex where flaring begins with the critical theoretical value
of $\dot m$ (Bildsten 1998) which demarcates the r\'egime of unstable He burning in a mixed H/He
environment from the r\'egime of stable H and He burning at higher values of $\dot m$. 
Bildsten's estimated uncertainties of $\pm$30\% are shown as dotted lines. It is clear
that in all three sources the measured $\dot m$ agrees with the critical value, and we
propose that as the sources descend the normal branch burning proceeds smoothly until at the
soft apex it becomes unstable at which point flaring immediately begins. During unstable burning
it is the emission of the neutron star that increases (Fig. 3 right, upper panel) providing 
the increase in total luminosity at constant $\dot M$. It can also be seen from Fig. 2 (left, lower panel)
that there may be an increase in blackbody radius in strong flaring as seen in GX\th 340+0 beyond the radius
of the neutron star, i.e. to 15 - 20 km. Such values have been obtained previously and may represent
an effect similar to radius expansion in the burst sources.

\section{Discussion}

We have shown that application of the extended ADC emission model for LMXB provides good
fits to the spectra of three Z-track sources at all positions along the Z-track. Morover,
the physical interpretation of the results is straightforward and strongly suggests an
explanation of the Z-track phenomenon, unlike use of the Eastern model in which interpretation
does not suggest a clear physical model. Our explanation of the Z-track is that the soft apex
is the lowest luminosity state of the source with minimum $\dot M$, with emission taking place
from the whole neutron star which has its lowest temperature.
On the normal branch, the increase of intensity and ADC luminosity
suggest an increase of $\dot M$ leading to a heating of the neutron star and a strong increase
in radiation pressure close to the neutron star. We suggest that this has a strong effect on the inner
accretion disk causing disruption of the disk. The horizontal effect will not directly
remove matter from the disk, but because the unperturbed height of the inner disk in LMXB at
these luminosities greater than 10$^{38}$ erg s$^{-1}$ is typically 50 km, the radiation pressure
can also act in a direction close to vertical blowing away material from the upper layers of the disk.
For the strongly super-Eddington fluxes that we measure close to the equatorial emitting zone of the
neutron star the effects can be very strong, and we propose that a substantial fraction of
the mass accretion rate flowing radially inwards in the disk is diverted vertically upwards
and is ejected from the system as massive blobs of plasma forming the jets above and below the disk.

Thus we propose that high radiation pressure is a {\it necessary} condition for jet formation
(but may not necessarily be a sufficient condition). The possible collimating effect of the
conical gaps in the inner accretion disk was previously suggested by Lynden-Bell (1978), and
the importance of radiation pressure in jet formation was discussed by 
Bisnovatyi-Kogan \& Blinnikov (1977) and Begelman \& Rees (1984).

The results also provides strong evidence that the flaring branch consists of unstable nuclear
burning and we obtain good agreement for the onset of the flaring branch with the theoretical
boundary between stable and unstable burning.

\section{Conclusions}

We show that the radiation pressure of the enitting part of the neutron star is very strong
at the hard apex and horizontal branch of the Z-track in three sources, exactly correlating with
the parts of the Z-track where radio emission is observed showing the presence of jets, and we
suggest that strong radiation pressure is a necessary condition for jet formation.

\begin{acknowledgements}
This work was supported by the UK Particle Physics and Astronomy Research Council (PPARC) and by 
the Polish Committee for Scientific Research (KBN) under grant no. KBN-1528/P03/2003/25.
\end{acknowledgements}


\begin{thebibliography}{99}

\bibitem[]{} Agrawal, V. K., Sreekumar, P., 2003, MNRAS, 346, 933

\bibitem[]{} Begelman, M. C., Rees, M. J., 1984, MNRAS, 206, 209

\bibitem[]{} Bildsten, L., 1998, in Proc NATO ASIC 515, The Many Faces of Neutron Stars, eds. R. Buccheri,
J. van Paradijs \& M. A. Alpar, Dordrecht-Kluwer, 419

\bibitem[]{} Bisnovatyi-Kogan G. S., Blinnikov S. I., 1977, A\&A, 59, 111

\bibitem[]{} Church, M. J., Ba\l uci\'nska-Church, M., 1995, A\&A, 300, 441

\bibitem[]{} Church, M. J., Ba\l uci\'nska-Church, M., 2004, MNRAS, 348, 955


\bibitem[]{} Church M. J., Halai G. S., Ba\l uci\'nska-Church M., 2006, A\&A, 460, 233

\bibitem[]{} di Salvo, T., Farinelli, R., Burderi, L., et al.,
2002, A\&A, 386, 535

\bibitem[]{} Done, C., \.Zycki, P., Smith, D. A., 2002, MNRAS, 331, 453

\bibitem[]{}
Fomalont, E. B., Geldzahler, B. J., Bradshaw, C. F., 2001, ApJ, 558, 283

\bibitem[]{}
Fujimoto, M. Y., Hanawa, T., Miyaji, S. 1981, ApJ, 247, 267

\bibitem[]{}
Fushiki, I., Lamb, D. Q. 1987, ApJ, 323, L55

\bibitem[]{} Hasinger, G., Priedhorsky, W., Middleditch, J, 1989, ApJ, 337, 843

\bibitem[]{}
Hasinger, G., van der Klis, M., Ebisawa, K., Dotani, T., \hbox{Mitsuda, K.} 1990,
A\&A, 235, 131

\bibitem[]{} Lynden-Bell, D., 1978, Phys Scripta 17, 185

\bibitem[]{} Psaltis, D., Lamb, F. K., Miller, G. S., 1995, ApJ, 454, L137

\bibitem[]{}
Priedhorsky, W., Hasinger, G., Lewin, W. H. G., et al., 1986, ApJ, 306, L91

\bibitem[]{}
Schatz, H., Bildsten, L., Cumming, A., Wiescher, M. 1999, ApJ, 524, 1014

\bibitem[]{} van der Klis, M., Stella, L., White, N. E., Jansen, F., Parmar, A. N.,
1987, ApJ, 316, 411

\end{thebibliography}
\end{document}